\documentstyle[twocolumn,aps,epsf]{revtex}

\newcommand{\te}{\tilde e}
\newcommand{\del}{{\bf\nabla}}
\def\gr{{g\over r}}
\def\at{{\pmb{\cal A}}}
\def\al{\tilde{\!\!{\pmb{\cal A}}}}
\def\s{\sin\theta}
\def\c{\cos\theta}
\def\ut{{\hat\theta}}
\def\uf{{\hat\phi}}
\def\ap{{\bf A}_+}
\def\am{{\bf A}_-}
\def\ans{{\bf A}_{NS}}
\def\f{\phi}
\def\pmb#1{\setbox0=\hbox{$#1$}%
\kern-.025em\copy0\kern-\wd0
\kern.05em\copy0\kern-\wd0
\kern-.025em\raise.0433em\box0 }

\begin{document}
\title { 
{\bf On the Dirac quantization condition.}}

\author{G.~I. Poulis$^1$ and P.J.~Mulders$^{1,2}$}
\address{$^1$National Institute for Nuclear Physics and High--Energy
Physics (NIKHEF)\\
P.O. Box 41882, NL-1009 DB Amsterdam, the Netherlands\\
and\\
$^2$Department of Physics and Astronomy, Free University \\
De Boelelaan 1081, NL-1081 HV Amsterdam, the Netherlands}
\maketitle

\begin{abstract}{\em }
\indent{\bf Abstract:}
We revisit the Dirac quantization condition
for string--like and string--less (but multi--valued)
magnetic monopole potentials. In doing so we  allow for an
{\it a priori} different coupling ${\tilde e}$
associated with the longitudinal components of the gauge potential.
By imposing physical criteria in the choice of
the longitudinal--transverse decomposition we show that
---in contrast to some recent claims--- the ``unphysical'' coupling
$\tilde e$ does not appear in the quantization condition.

\end{abstract}

\medskip
\mbox{}
\medskip

\noindent As first discussed by Dirac in his seminal paper~\cite{Dirac}
the existence of magnetic monopoles with magnetic charge $g$
implies quantization of the electric charge $e$ according to
\begin{equation}\label{ena}
e g = n/2,\quad n\in Z \ .
\end{equation}
The consistency of $U(1)$ gauge theory in the presence of
such magnetic monopoles has been recently questioned by
He, Qiu and Tze~\cite{He,VPI_14,Add} who have proposed a
generalized formulation of QED where they allow for two
different coupling constants $e$ and $\te$ associated, respectively,
with the transverse (physical) and longitudinal (unphysical) components
of the gauge field. By considering both string--type and string--less
(see below) monopole potentials they have argued that the conventional
quantization condition, Eq.~(\ref{ena}), is replaced by one where the
unphysical coupling, $\te$, enters, {\it i.e.},
\begin{equation}\label{dyo}
\te g = n/2, \quad n\in Z.
\end{equation}
\noindent
Thus, they conclude, since in the above equation a physical
coupling, $g$, is constrained by an unphysical one, $\te$,
the only viable scenario is that the
monopole charge $g$ has to be zero, unless one enforces
$e=\te$ which they view as unacceptable since no physical process
would ever ``see'' $\te$. However, it is very hard to reconcile such
a statement with the prolific work on monopoles in
lattice--regularized field theory: monopoles have been shown
not only to be present but also
to drive confinement in the strong coupling regime, both
analytically~\cite{anal} and in
Monte Carlo simulations~\cite{latt} and their non--observation
is understood as a {\it dynamical} effect, namely,
exponential vanishing of the monopole density
as one approaches the continuum limit, whereas the arguments
quoted by the authors
are essentially {\it symmetry} arguments that would apply to
both strong and weak couplings. Moreover, it is not clear whether
one can call $\te$ ``unphysical''
for processes involving {\it virtual} photons (electron--hadron
scattering, $e^+\!-\!e^-$ annihilation {\it etc.}) where the
longitudinal components of the gauge field (and therefore $\te$
itself) do contribute to the S--matrix.
Notwithstanding this observation we shall leave it aside and
restrict ourselves to the discussion of {\it how to properly
derive the quantization condition in this
generalized version of U(1) gauge theory}.  In a first attempt
to examine the arguments of the above authors we have pointed
out~\cite{us} that arriving to Eq.~(\ref{dyo}) instead of the
conventional quantization condition, Eq.~(\ref{ena}),
depends crucially on the way the decomposition of the gauge field in
longitudinal and transverse pieces is carried out. In particular, we
noticed that {\it the Dirac string--type solutions are
always divergenceless}.  Thus, one could take them to be purely
transverse and therefore there would be no term coupled to the
unphysical coupling $\te$: this coupling (whether unphysical or not)
need not enter the quantization condition. However, the property of
being divergenceless is not true for
string--less, multi--valued potentials which provide
a monopole field as well (see below).
In this note we wish to examine all these types of monopole potentials
in a unified framework.  As a warm--up we will rederive the
quantization condition for the following three (static) monopole
magnetic potentials:
\begin{eqnarray}\label{tri}
{\bf A}_{\pm} &=& -\gr\left(\c\pm 1\over \s\right)\uf \\
\ans           &=& -\gr\s\f\ut \nonumber \ .
\end{eqnarray}
\noindent
The first two correspond to potentials with Dirac
strings in the $\pm\hat z$ axis.
Although initially formulated in terms of potentials with string--type
singularities, magnetic monopole fields can be generated by
non--singular, multi--valued magnetic potentials. The most well
known type of those  is the Wu and Yang construction~\cite{Wu},
where one defines a {\it locally} non--singular potential
which is gauge equivalent to a string--type one~\cite{lump}.
A less familiar monopole potential is the above $\ans$
potential~\cite{ans} which is explicitly {\it multi--valued} due
to the presence of $\phi$, but {\it non--singular} (except at
the origin of course) and is again related via
a gauge transformation~\cite{VPI_14} to the string--type potentials.
\noindent
We derive a quantization condition (amongst other ways, see for
example~\cite{CL}) by requiring that the phase factor for
a charged particle's wavefunction, is unobservable, that is
\begin{equation}\label{extra}
\exp \{ e\oint_{\Gamma}{\bf A}\cdot{\bf dl}\} =\exp(2 i n\pi) =
 1, n\in Z \ ,
\end{equation}
where $\Gamma$ is {\it any} closed loop shrunk to zero (that is,
surrounding a minimal surface with zero area).
\begin{figure}[htb]
\begin{center}
\leavevmode
\epsfxsize=6.5cm\epsfbox{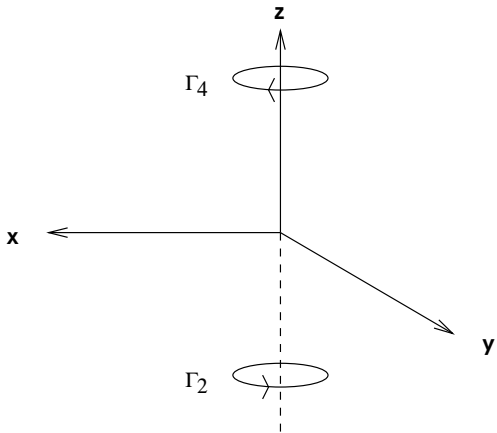}
\caption{The $\Gamma_2$ and $\Gamma_4$ contours (see text).}
\end{center}
\label{fig:dtria}
\end{figure}
\noindent
For the ${\bf A}_+$ potential we choose a loop $\Gamma_4$
in the $-\uf$ direction (clockwise) with $r$ fixed and $\theta
\rightarrow 0$ (Fig. 1) and obtain $4\pi e g = 2\pi n$, which is
the Dirac quantization condition, Eq.~(\ref{ena}). Similarly,
for $\bf A_-$ we use a (counterclockwise)
loop with $\theta\rightarrow \pi$ (labeled $\Gamma_2$ in Fig. 1)
and obtain the same condition. The nontrivial content of this
constraint is that the loop has been chosen so as to encircle
the singular string: thus, Stokes theorem
(which would state that the flux is zero since the encompassing
area is zero)
 is not applicable. Had we chosen loop $\Gamma_2$ for the potential
$\bf A_+$, the phase factor would be {\it trivially} unity,
and no quantization condition could (or should) be obtained from this
(poor) choice of integration contour. This is somehow an obvious
statement
but it will be useful later on in our discussion.
\begin{figure}[htb]
\begin{center}
\leavevmode
\epsfxsize=6.5cm\epsfbox{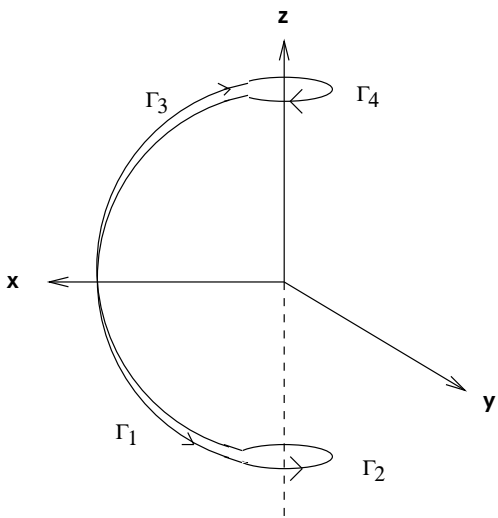}
\caption{The $\Gamma=\Gamma_1+\Gamma_2+\Gamma_3+\Gamma_4$ contour (see text).}
\end{center}
\label{fig:dtrib}
\end{figure}
\noindent
Finally, for the $\ans$ potential we choose the following
closed loop $\Gamma$ (Fig. 2):
start at $(r,\theta,\phi)=(R,0,0)$ then go to  $(R,\pi,0)$
along the $\ut$ direction ($\Gamma_1$), then to $(R,\pi,2\pi)$ along
$\uf$ ($\Gamma_2$ above), then to $(R,0,2\pi)$ along $-\ut$ ($\Gamma_3$),
and finally clockwise (along $-\uf$) back to $(R,0,0)$ ($\Gamma_4$).
The only nonvanishing contribution comes from $\Gamma_3$ and equals
$2\pi e g\int_0^{\pi}\s d\theta=4\pi e g$ and we arrive again at the Dirac
quantization condition, Eq.~(\ref{ena}).
We note that the loop $\Gamma$ constructed in this way, has several
nice features. For single-valued potentials it reduces to
$\Gamma_2 + \Gamma_4$, which depending on where one has a singularity
reduces to just $\Gamma_2$ or $\Gamma_4$. For multi-valued
potentials it picks up contributions through $\Gamma_1 + \Gamma_3$.

Let us now introduce the ``generalized'' covariant derivative used
in Ref.~\cite{He}
	\begin{equation}\label{ndyo}
	D_{\mu}=\partial_{\mu} - i e {\cal A}_{\mu} -
	i \tilde e \tilde {\cal A}_{\mu} \ .
	\end{equation}
Here the gauge field $A_{\mu}$ is decomposed into transverse,
 ${\cal A}_{\mu}=T_{\mu\nu}A^{\nu}$,
and longitudinal, $\tilde{\cal A}_{\mu}=L_{\mu\nu}A^{\nu}$, components,
coupled to charges $e$ and $\te$, respectively;
we employ the projectors $L_{\mu\nu}=\partial_{\mu}\partial_{\nu}/
\partial^2$ and $T_{\mu\nu}=g_{\mu\nu}-L_{\mu\nu}$. The longitudinal
components do not enter the field strength tensor $F_{\mu\nu}$ and are
unphysical. This theory is invariant under local $U(1)$
transformations
	\begin{equation}\label{tria}
	e{\cal A}_{\mu}(x) + \te\tilde{\cal A}_{\mu}(x)
	\rightarrow e{\cal A}_{\mu}(x) + \te\tilde{\cal A}_{\mu}(x)
	 - \partial_{\mu} \Omega(x) \ .
	\end{equation}
By applying the projectors $L_{\mu\nu},
\; T_{\mu\nu}$ on both sides of (\ref{tria}) one obtains
	\begin{eqnarray}\label{trib}
	{\cal A}_{\mu}(x) &\rightarrow & {\cal A}_{\mu}(x)
	- {1\over e}\partial_{\mu}\Omega(x) + {1\over e}
	\partial_{\mu}{1\over\partial^2}\partial^2\Omega(x)\nonumber\\
	\tilde{\cal A}_{\mu}(x)
	&\rightarrow & \tilde{\cal A}_{\mu}(x)
 	- {1\over\te} \partial_{\mu}{1\over\partial^2}\partial^2
	 \Omega(x) \ .
	\end{eqnarray}
Thus, in the static case, as noticed in~\cite{He},
 the transverse components are
left invariant and only the longitudinal ones change:
	\begin{eqnarray}\label{tric}
	{\bf\cal A}(\bf r) &\rightarrow & {\bf\cal A}(\bf r)
	\nonumber\\
	\tilde{\bf\cal A}({\bf r})
	&\rightarrow & \tilde{\bf\cal A}({\bf r})
	 - {1\over\te} {\bf\nabla}\Omega({\bf r}) \ ,
	\end{eqnarray}
However, as pointed out in~\cite{us} this statement is
ambiguous when the potential $\bf A$ is divergenceless, for then
$\partial^{-2}A_{\mu}$ cannot be defined. In fact, this is
the case for {\it all} string--type potentials, since~\cite{VPI_14,us}
	\begin{equation}\label{tes}
	{\pmb\nabla}\cdot{\bf  A}=0,\>\mbox{with}\>
	{\bf A}({\bf r})  = - \int_{\Gamma} {\bf du}
	\times {\bf B}({\bf r} - {\bf u}) \ .
	\end{equation}
Here ${\bf B(\bf x)}= - g {\pmb\nabla} ({1\over 4\pi x})$
and the Dirac string lies along
a generic single--valued semi--infinite path $\Gamma$.
 But if the potential is divergenceless we can as well take the
longitudinal part to be zero. Moreover, if we consider a (static)
transformation $\Omega_{\Gamma,\Gamma'}$ that moves the Dirac string
to lie along a different path $\Gamma'$ the above property
guarantees that $\partial^2\Omega_{\Gamma,\Gamma'}=0$
 and that  under this gauge transformation the longitudinal component
remains the same (trivially, since it is zero) while the transverse changes
	\begin{eqnarray}\label{trid}
	{\bf\cal A}({\bf r}) &\rightarrow& {\bf\cal A}({\bf r})
	- {1\over e} {\bf\nabla}\Omega({\bf r})
	\nonumber\\
	\tilde{\bf\cal A}({\bf r})
	&\rightarrow & \tilde{\bf\cal A}({\bf r}) = 0 \ \ .
	\end{eqnarray}
This equation should be contrasted with Eq.~(\ref{tric}).
In order to be more specific let us discuss the
longitudinal--transverse decomposition for the potentials in Eq.~(\ref{tri}).
He, Qiu and Tze offer the following decomposition into transverse and
longitudinal components:
	\begin{equation}\label{pente}
	{\cal A}_{\mp}^{(\rm{I})}
	= -\gr {\c\over\s}\uf,\;\;
	\tilde{\cal A}_{\mp}^{(\rm{I})}
 	= \pm\gr{1\over \s}\uf\ .
	\end{equation}
Thus, the transverse piece is the same for the two strings and the
gauge transformation that maps one string solution to the other
is of the type (\ref{tric}) with $\Omega({\bf r}) = 2 \te g \phi$.
However, as we said above, $\del^2\Omega=0$ and thus we could as well
have the decomposition
	\begin{equation}\label{dyo1}
	{\cal A}_{\mp}^{(\rm{II})} = -\gr {\c\mp 1\over\s}\uf,\;\;
	\tilde{\cal A}_{\mp}^{(\rm{II})} = 0\nonumber \ .
	\end{equation}
Notice that the  gauge transformation that connects the two strings  is
now of the type  (\ref{trid}) with $\Omega({\bf r}) = 2 e g \phi$
which is the same as above but with the physical coupling, $e$, appearing
 instead of the unphysical one, $\te$.
What about the string--less potential $\ans$?
 In this case we cannot take
it to be purely transverse, since
	\begin{equation}\label{exi}
	\del\cdot{\s\f\over r}\ut = 2{\cos\theta\over r^2} \phi \ne 0 \ .
	\end{equation}
It is straightforward to check that all of the following are
legitimate longitudinal--transverse decompositions for $\ans$:
	\begin{eqnarray}\label{dyo2}
	{\cal A}_{NS}^{(\rm{I})}
	 = -\gr {\c\over\s}\uf, &\;\; &
	\tilde{\cal A}_{NS}^{(\rm{I})}
 	= \gr {\c\over\s}\uf -\gr\s\f\ut\\
	{\cal A}_{NS}^{(\rm{II})} = -\gr {\c\pm 1\over\s}\uf,&\;\; &
	\tilde{\cal A}_{NS}^{(\rm{II})} = \gr {\c\pm 1\over\s}\uf -\gr\s\f\ut
\nonumber\ .
\end{eqnarray}
As before, decomposition (I) is the one used by He, Qiu and Tze~\cite{VPI_14}.
In the decompositions (II+,II-) we have used our experience from the
string--type potentials above to move $\pmb\nabla\phi=(1/ r\s)\uf$
pieces between the longitudinal
and transverse parts since these terms can be equally well considered
to be either longitudinal or transverse.\footnote{they have both zero curl
and zero divergence}

In the next step, we are going to calculate the flux through the various
loops that we have employed before, separating the flux coming from the
longitudinal and the transverse parts.
In order to discuss the results we will consider three
criteria, starting with the one in (\ref{extra})
for this generalized version of QED:
\begin{itemize}
\item{\underline{criterion (A)}:}
{ the flux through any closed loop shrunk to zero should
be unobservable}, that is
\begin{equation}\label{nena}
\Bigl\{ e\oint_{\Gamma}\at\cdot{\bf dl} +
               \tilde e\oint_{\Gamma} \al \cdot {\bf dl} \Bigr\} =
2\pi n,\; n\in Z \ ,
\end{equation}
where, as before, $\Gamma$ is a closed loop
surrounding a minimal surface with zero area.
\end{itemize}
We wish to emphasize that a finite contribution to the left
 hand side of Eq.~(\ref{nena})
for ``zero area'' loops $\Gamma$ can arise either because the potential is
{\it singular} inside $\Gamma$ or because there is no singularity but the
potential is
{\it multi--valued} around $\Gamma$. Thus, we impose two more  criteria:
\begin{itemize}
\item{\underline{criterion (B)}:}
	we should not expect to	obtain a constraint in the form of a
        quantization condition for loops inside which
	the potential is both {\it smooth} and {\it single--valued}.
\end{itemize}
This criterion should be imposed  because in the region where the potential
is smooth and  single--valued it is irrelevant whether it arises from a
monopole string (which we want to make unobservable) or from a semi--infinite
solenoid (which is put there by hand and is certainly observable).
Any condition in this region would therefore constraing the possible values
that the flux through the solenoid can take which is unphysical.

\begin{itemize}
\item{\underline{criterion (C)}:}
	When we decompose the string--less potential $\ans$ into longitudinal
	and transverse parts we should be careful to avoid ---if possible---
	introducing singularities because then we undo the very
        motivation for introducing such potentials,
	namely, that they are non--singular!
\end{itemize}
 It is easy to show that this latter criterion (C)
can be {\it locally} (that is, for a subset of $R^3$)
satisfied: just add enough $\pmb\nabla\phi= (1/r\s)\uf$ terms so
as to compensate for the singularity. For example, a singularity of the type
$(\c/ r\s)\uf$ which extends over the {\it whole} $z$ axis can,
by adding ${\pm\pmb\nabla\f}$, be reduced to the {\it half} $z$ axis.
That means that we introduce a decomposition that
is different in various space regions (\`a la Wu and Yang),
but the potential is multivalued anyway so that's not a problem.

In Table 1 we show the results for the $\Gamma_4$ loop.
We have made the choice ${\cal A}_{NS}={\cal A}_-$ for loop $\Gamma_4$,
so as to keep the decomposition non--singular,
in accordance with criterion (C).

\renewcommand{\arraystretch}{1.5}

\begin{center}
\begin{tabular}{|c|c|c||c|c|}\hline
\multicolumn{5}{|c|}
{Table 1: the $r$=fixed, $\theta\rightarrow 0$ loop $\Gamma_4$} \\ \hline\hline
& \multicolumn{2}{c||} {decomposition (I)} &  \multicolumn{2}{c|}
{decomposition (II)} \\ \cline{2-5}
& {$e\oint_{\Gamma_4}\at\cdot{\bf dl} $}
& {$\tilde e\oint_{\Gamma_4}\al\cdot {\bf dl} $}
& {$e\oint_{\Gamma_4}\at\cdot{\bf dl} $}
& {$\tilde e\oint_{\Gamma_4}\al\cdot {\bf dl} $} \\ \hline
 $\ap$  & $+2\pi e g$ & $+2\pi\tilde eg$ & $+4\pi eg$  &  $   0  $ \\ \hline
 $\am$  & $+2\pi e g$ & $-2\pi\tilde e g$ & $  0    $  &  $   0   $ \\ \hline
 $\ans$ & $+2\pi e g$ & $-2\pi\tilde e g$ & $  0    $  &  $   0   $ \\ \hline
\end{tabular}
\end{center}
Criterion (A) then implies the following constraints
for the loop $\Gamma_4$ and decomposition (I) for the three
potentials ($n\in Z$ throughout):
	\begin{eqnarray}\label{alex}
 	\ap&: &  g(e+\tilde e) = n
		\Rightarrow \left\{ \begin{array}{ll}        n=g=0, &
 		\mbox{if $\tilde e$ arbitrary}\\
        	eg=n/2, & \mbox{if $\tilde e=e$}     \end{array} \right.
 		\\
	\am &: &  g(e-\tilde e) = n
		\Rightarrow \left\{ \begin{array}{ll}
        	n=g=0, & \mbox{if $\tilde e$ arbitrary}\\
       		\mbox{no constraint}, & \mbox{if $\tilde e=e$}
       		\end{array}\right.  \nonumber \\
   	\ans &: &  g(e-\tilde e) = n
 		\Rightarrow \left\{ \begin{array}{ll}
		n=g=0, & \mbox{if $\tilde e$ arbitrary}\\
        	\mbox{no constraint}, & \mbox{if $\tilde e=e$ \ ,}
       		\end{array} \right.  \nonumber
	\end{eqnarray}
while for the same loop $\Gamma_4$ decomposition (II) implies
\begin{eqnarray}\label{pen}
 \mbox{$\ap$}&:&  ge = n/2 \\
 \mbox{$\am$}&:&  \hbox{no constraint}\nonumber \\
 \mbox{$\ans$}&:&  \hbox{no constraint}\nonumber \  .
\end{eqnarray}
A number of observations can be made here:
\begin{enumerate}
\item{} Decomposition (II) meets all criteria (A), (B) and (C). Moreover,
        the {\it only} constraint it leads to is the conventional quantization
        condition  $ge=n/2$.
\item{} Decomposition (I) implies {\it individual} quantization
        conditions for $e\!-\!\te$ and $e\!+\!\te$. If we treat
        $\te$ as independent of $e$ and subtract these constraints
        we get Eq.~(\ref{dyo})
        used in~\cite{He,VPI_14} to prove that QED is inconsistent
        with monopoles. However, we view this result as unphysical
        since for $\te$ arbitrary decomposition (I) violates {\it both}
        (B) and (C) criteria.
        In particular it violates (B) by  leading to a constraint
        (the second of Eqs.~(\ref{alex}) above) coming from a potential, $\am$,
        which is smooth and single--valued in the area of the loop $\Gamma_4$.
        It also violates (C) by introducing singularities in the vicinity of
        loop $\Gamma_4$ for the string--less potential $\ans$.
\end{enumerate}

Analogous remarks can be made for the $\ap$ potential in the case of
the $\Gamma_2$ loop (see Table 2).
In this case we have chosen ${\cal A}_{NS}={\cal A}_+$, in accordance with
criterion (C).
\begin{center}
\begin{tabular}{|c|c|c||c|c|}\hline
\multicolumn{5}{|c|}
{Table 2: the $r$=fixed, $\theta\rightarrow \pi$ loop $\Gamma_2$} \\
\hline\hline
& \multicolumn{2}{c||} {decomposition (I)} &  \multicolumn{2}{c|}
{decomposition (II)} \\ \cline{2-5}
& {$e\oint_{\Gamma_2}\at\cdot{\bf dl} $}
& {$\tilde e\oint_{\Gamma_2}\al\cdot {\bf dl} $}
& {$e\oint_{\Gamma_2}\at\cdot{\bf dl} $}
& {$\tilde e\oint_{\Gamma_2}\al\cdot {\bf dl} $} \\ \hline
 $\ap$  & $+2\pi e g$ & $-2\pi\tilde e g$ & $   0     $  &  $   0      $ \\
\hline
 $\am$  & $+2\pi e g$ & $+2\pi\tilde e g$ & $+4\pi e g$  &  $   0      $ \\
\hline
 $\ans$ & $+2\pi e g$ & $-2\pi\tilde e g$ & $   0     $  &  $   0      $ \\
\hline\end{tabular}
\end{center}
Notice that in our decomposition (II) so far these loops do not lead to a
quantization condition stemming from the string--less, but
multi-valued potential $\ans$. However, also for ordinary QED ($e$ =
$\tilde e$)
one did not obtain such a condition from these loops. So let's discuss the
results for the loop $\Gamma$ (Fig. 2).
We have to add the contributions of
Tables 1 and 2 and also add the contribution of the $\ut$ part of $\ans$
to $\Gamma_3$ (the contribution to $\Gamma_1$ vanishes because $\phi = 0$).
The results are presented in Table 3. One sees that (a) the {\it only}
quantization condition that can be obtained in this case is the conventional
one $e g = n/2$, {\it independent of the transverse--longitudinal
decomposition one uses} and (b) that the longitudinal terms lead to no
constraint whatsoever.
This is a direct consequence of the fact that the piece $\pmb\nabla\phi$
does not contribute to the flux through the loop $\Gamma$ in Fig. 2. Therefore,
irrespective of the decomposition of the potential, the condition obtained
from this loop is the ordinary quantization condition in Eq.~(\ref{dyo}).
\begin{center}
\begin{tabular}{|c|c|c||c|c|}\hline
\multicolumn{5}{|c|}
{Table 3: the loop $\Gamma=\Gamma_1+\Gamma_2+\Gamma_3+\Gamma_4$}\\ \hline\hline
& \multicolumn{2}{c||} {decomposition (I)} &  \multicolumn{2}{c|}
{decomposition (II)} \\ \cline{2-5}
& {$e\oint_{\Gamma}\at\cdot{\bf dl} $}
& {$\tilde e\oint_{\Gamma}\al\cdot {\bf dl} $}
& {$e\oint_{\Gamma}\at\cdot{\bf dl} $}
& {$\tilde e\oint_{\Gamma}\al\cdot {\bf dl} $} \\ \hline
 $\ap$  & $+4\pi e g$ & $   0     $ & $+4\pi e g$   &  $   0      $ \\ \hline
 $\am$  & $+4\pi e g$ & $   0     $ & $+4\pi e g$   &  $   0      $ \\ \hline
 $\ans$ & $+4\pi e g$ & $   0     $ & $+4\pi e g$   &  $   0      $ \\ \hline
\end{tabular}
\end{center}
We summarize: the fact that criterion (A) is applicable to {\it any} closed
loop
allow us to recover the conventional quantization condition $ge=n/2$
not only for string--type potentials where their divergenceless makes this
result
 quite obvious {\it but also when considering the string--less potential $\ans$
which is not purely transverse}. This result stems from the
decomposition--independent
$\Gamma$ contour integrations  in Table 3.  Moreover, by imposing some physical
 criteria (B) and (C) for choosing a physical
longitudinal--transverse decomposition we have shown that the  freedom
to move $\pmb\nabla\phi$ parts between longitudinal
and transverse pieces amounts to the following:
\begin{itemize}
\item{} The {\it only} quantization condition that can be obtained is the
        conventional one, Eq.~(\ref{ena}).
\item{} No constraint on $\tilde e$ is imposed.
\end{itemize}
Thus, even by treating the longitudinal coupling $\te$ as arbitrary
(despite the
questions this raises for virtual photons) we have shown that
quantum electrodynamics is consistent with magnetic monopoles.

\bigskip
{\bf Acknowledgements:}
This work is supported in part by the Foundation for Fundamental
Research on Matter (FOM) and the National organization for Scientific Research
(NWO) as well as the Human Capital and Mobility
Fellowship ERBCHBICT941430.


\begin{references}
\bibitem{Dirac}  P.A.M. Dirac, Proc. Poy. Soc. (London)
                 Ser. A, {\bf 133}, 60 (1931).
\bibitem{He}     H.-J. He, Z. Qiu and C.-H. Tze, Z. Phys.  C {\bf 65}, 175
(1994).
\bibitem{VPI_14} H.-J. He, Z. Qiu and C.-He, Virginia Polytechnic Institute
preprint
                 VPI-HEP-93-14 (1993).
\bibitem{Add}    H.-J. He and C.-H. Tze,
                 Addendum to ``Inconsistency of QED in the presence of
                 Dirac Monopoles'', Z. Physik C (in press).
\bibitem{us}     G.I. Poulis and P.J. Mulders, NIKHEF preprint
                 {\it NIKHEF-94-P11}, Z. Physik C (in press).
\bibitem{anal}   T. Banks, R. Myerson and J. Kogut, Nucl. Phys. B {\bf 129} 493
(1977);
		 A.M. Polyakov in ``Gauge Fields and Strings'', Harwood Academic Publishers,
                 Chur, Switzerland (1987).
\bibitem{latt}   T.A. DeGrand and D. Toussaint, Phys. Rev. D {\bf 22},
                 2478 (1980), H.~D.~Trottier and R.~M.~Woloshyn, Phys. Rev. D
                 {\bf 48}, 4450 (1993).


\bibitem{Wu}	 T.T. Wu and C.N. Yang, Phys. Rev. Lett. {\bf 13}, 380 (1964).
\bibitem{lump}   Sidney Coleman, Appendix 5 in ``Classical lumps and their
quantum
                 descendants'', page 260, Aspects of Symmetry, Cambridge
University
                 Press (1988).
\bibitem{ans}    H.A. Cohen, Progr. Theor. Phys. Vol 50, No. 2, (1973).
\bibitem{CL}     T.-P. Cheng and L.-F. Li in ``Gauge Theory of Elementary
Particle
                 Physics'', Oxford University Press (1984).

\end{references}
\end{document}